\documentclass[numberedappendix,apj]{emulateapj-rtx4}

\def\heao{{\it HEAO 1\/}}

\def\ltsima{$\; \buildrel < \over \sim \;$}
\def\simlt{\lower.5ex\hbox{\ltsima}}
\def\gtsima{$\; \buildrel > \over \sim \;$}
\def\simgt{\lower.5ex\hbox{\gtsima}}
\def\kms{\ifmmode{~{\rm km~s^{-1}}}\else{~km s$^{-1}$}\fi}
\def\lsim{\lower0.3em\hbox{$\,\buildrel <\over\sim\,$}}
\def\gsim{\lower0.3em\hbox{$\,\buildrel >\over\sim\,$}}

\def\h2{H$_2$}

\def\heao1{{\it HEAO-1\/}}

\begin{document}
\title{Can we reproduce the X-ray background spectral shape using local AGN?}
\author{Ranjan V. Vasudevan\altaffilmark{1,*}, Richard F. Mushotzky\altaffilmark{1}, Poshak Gandhi\altaffilmark{2},\altaffilmark{3}}
\altaffiltext{1}{Department of Astronomy, University of Maryland, College Park, MD, 20742}
\altaffiltext{2}{Institute of Space and Astronautical Science, Japan Aerospace Exploration Agency, 3-1-1 Yoshinodai, chuo-ku, Sagamihara, Kanagawa 252-5210, Japan}
\altaffiltext{3}{Department of Physics, Durham University, South Road, Durham DH1 3LE, UK}
\altaffiltext{*}{ranjan@astro.umd.edu}

\begin{abstract}

The X-ray background (XRB) is due to the aggregate of active galactic nuclei (AGN), which peak in activity at $z \sim 1$ and is often modeled as the sum of different proportions of unabsorbed, moderately- and heavily-absorbed AGN.  We present the summed spectrum of a complete sample of local AGN (the Northern Galactic Cap of the 58-month Swift/BAT catalog, $z<0.2$) using 0.4--200~keV data and directly determine the different proportions of unabsorbed, moderately and heavily-absorbed AGN that make up the summed spectrum.  This stacked low redshift AGN spectrum is remarkably similar in shape to the XRB spectrum (when shifted to $z \sim 1$), but the observed proportions of different absorption populations differ from most XRB synthesis models.  AGN with Compton-thick absorption account for only $\sim 12\%$ of the sample, but produce a significant contribution to the overall spectrum.  We confirm that Compton reflection is more prominent in moderately-absorbed AGN and that the photon index differs intrinsically between unabsorbed and absorbed AGN.  The AGN in our sample account for only $\sim 1 \%$ of the XRB intensity.  The reproduction of the XRB spectral shape suggests that strong evolution in individual AGN properties is not required between $z \sim 0$ and $1$.

\end{abstract}

\section{Introduction}

The X-ray background (XRB) has been extensively studied from 1--500 keV and is the combined emission from AGN with varying degrees of obscuration and Compton reflection modifying the spectral shape of each object (e.g., \citealt{2009ApJ...696..110T}, \citealt{2007A&A...463...79G} - T09 and G07 respectively, hereafter). The excellent angular resolution and low background of \emph{Chandra} have allowed an analysis of  the properties of the individual AGN that make up most of the 1-6 keV XRB \citep{2006MNRAS.368.1735W}.  However, these objects lie at redshifts $\langle z \rangle \sim 1.0$ (e.g., \citealt{2008A&A...490.1039S}), and therefore the spectral quality is poor in comparison to nearby AGN. Additionally, spectral data are typically only available in the Chandra/XMM band (observed frame $<10$~keV) and do not extend to 15--40~keV, where the majority of the XRB power lies. It was therefore not possible to uniquely reconstruct the XRB spectrum across 1--500~keV by combining the emission from individual AGN. 

Numerous studies predict the different populations of AGN required to reproduce the XRB spectral shape, by simulating the observed XRB as the sum of different proportions of unabsorbed, absorbed Compton-thin and absorbed Compton-thick spectra, (G07, T09, \citealt{2012A&A...546A..98A}).  The fraction of Compton-thick sources remains uncertain as does the distribution of Compton reflection strengths and high energy cut-offs.  Recently \cite{2012A&A...546A..98A} showed that a wide range of these parameters can fit the observed XRB spectrum and redshift distribution of sources.  Without stricter constraints on the source properties, there remains a large uncertainty in the models.    We adopt a `bottom-up' approach to understanding the make-up of the XRB by examining the composite spectra of local AGN over 0.5--200~keV.  This approach was pioneered by \cite{2008A&A...482..517S} with a very thorough analysis of INTEGRAL and RXTE data (3--300~keV), and developed further by \cite{2011A&A...532A.102R} (10--300~keV). We utilize data from ASCA, \emph{Swift}/XRT, XMM and \emph{Swift}/BAT, and compare the composite with the XRB spectrum and synthesis models.  We show that the sum of the local sources provides an excellent match to the observed XRB shifted to $z\sim 0$ indicating that despite expected evolution in AGN population properties (due to observed changes in luminosity, accretion rate, host galaxy type, obscured AGN fraction between $z=1$ and now, e.g. \citealt{2005AJ....129..578B,2012ApJ...746...90A,2010A&A...509A..78D}), the overall combined emission from such sources stays the same.

\section{Sample}
The 58-month Swift Burst Alert Telescope (BAT) all-sky, hard-X-ray--selected survey of local AGN\footnote{http://heasarc.gsfc.nasa.gov/docs/swift/results/bs58mon/} ($\langle z \rangle = 0.043$) is complete and relatively unbiased to absorption, thanks to the efficient hard X-ray selection. We employ the detailed X-ray analysis of our companion paper \cite{2013ApJ...763..111V} (V13 hereafter) for a complete subsample from the 58-month catalog, the Northern Galactic Cap (all radio-quiet sources with Galactic latitude $b > 50^{\circ}$, flux limit $4 \times 10^{-12} \thinspace \rm erg \thinspace s^{-1} \thinspace cm^{-2}$). In V13, 0.4--10~keV data from XMM, \emph{Swift}/XRT and ASCA, along with Swift/BAT data (14--200~keV), are fitted to produce estimates of the column density, spectral slope and reflection fraction ($R$) for each object.

\section{Generating the stacked spectrum of the BAT AGN}
The V13 study provides a best-fitting model for each source. The data in the 0.4--10~keV band are from different satellites and of variable quality, so we use the best-fit models  to parameterize their spectra.  For each object, we extract the model using 1000 bins over 0.4--10~keV (using the `dummyrsp' command in {\sc xspec}) providing a spectrum for each object over a common energy grid. We sum flux densities over this grid to generate the summed spectrum. We perform this for the whole sample (95 objects with good spectra) and for different absorption regimes: unabsorbed ($\rm log \thinspace N_{\rm H} \thinspace < 21$, 27 objects), absorbed Compton-thin ($21 < \rm log \thinspace N_{\rm H} \thinspace < 24$, 51 objects) and absorbed Compton-thick ($\rm log \thinspace N_{\rm H} \thinspace > 24$, 12 objects), using the same $\rm log \thinspace N_{\rm H}$ bins as G07 for comparison with their results.

For thirteen objects, their $N_{\rm H}$ values are uncertain due to poorer quality 0.4--10~keV data.  We therefore produce two sets of summed spectra; one in which these objects take their minimal $N_{\rm H}$ values, and one in which they take their maximal $N_{\rm H}$ values.

\begin{figure*}
\centerline{
\includegraphics[width=9.0cm]{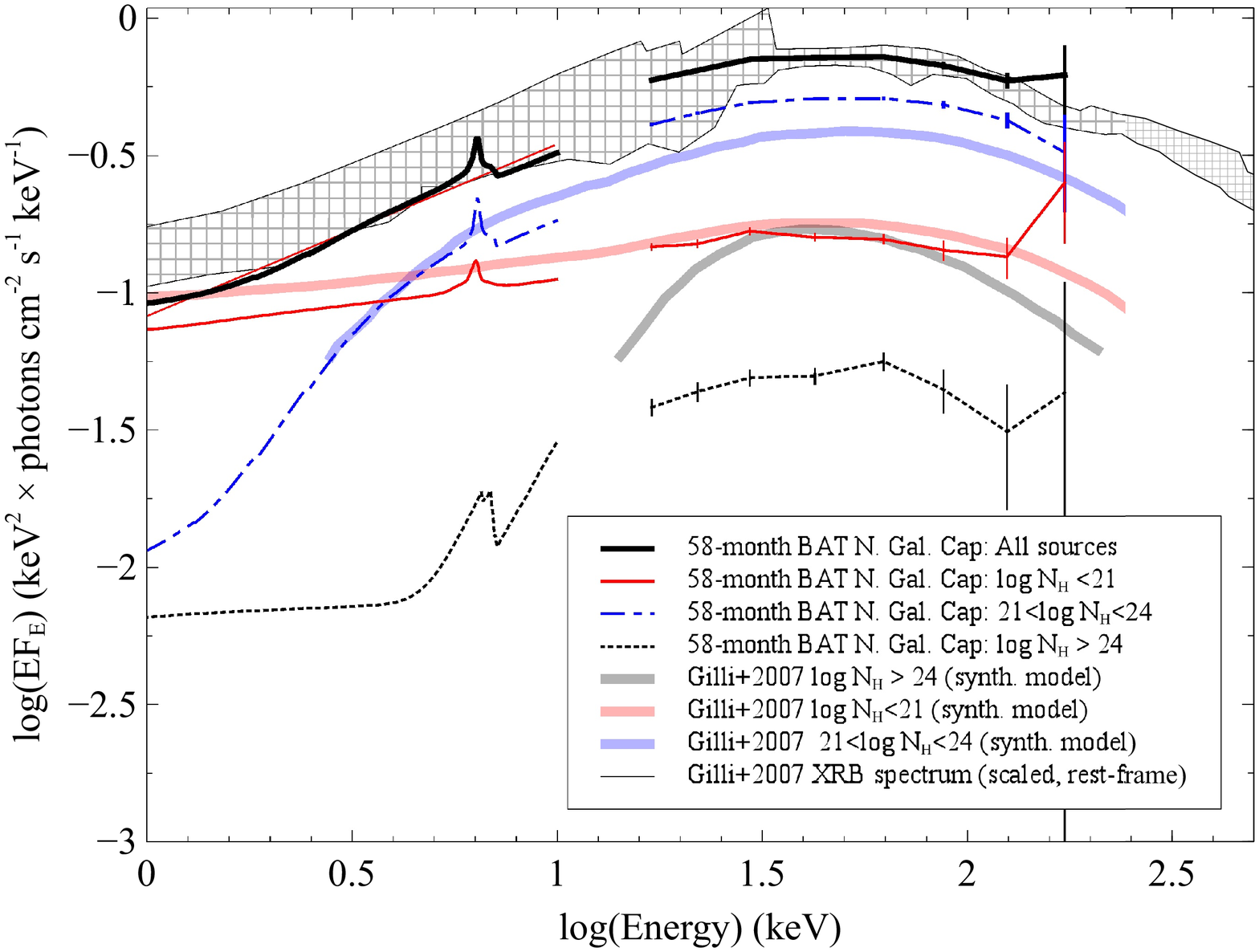}\\\\
\includegraphics[width=9.0cm]{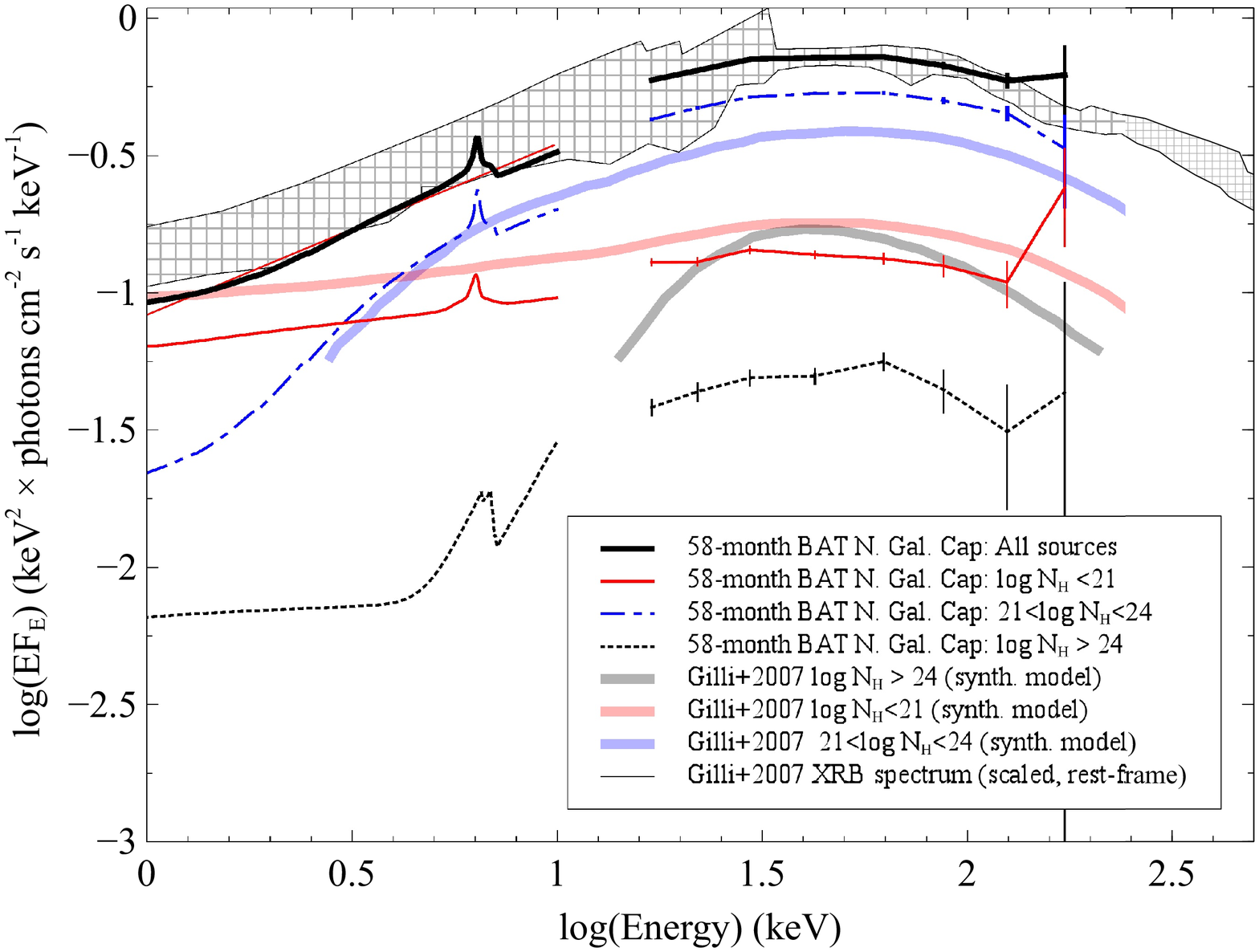}
}
\caption{Stacked 0.4--200~keV spectrum for the sources in our sample, comparing with G07 XRB synthesis model.  \emph{Left panel}: ambiguous $N_{\rm H}$ sources take their minimal $N_{\rm H}$. \emph{Right panel}: such sources have their maximal $N_{\rm H}$.  See inset key for details. \label{stackedspectra}}
\end{figure*}

Above 10~keV, the eight energy channels of BAT (at fixed energies between 14--200~keV), already provide a common energy grid over which to sum the fluxes.  We generate the summed  BAT spectra as above, for the whole sample and for different absorption regimes. For comparison with previous studies, we convert the stacked spectra to $E F_{E}$ format (Fig.~\ref{stackedspectra}).  

The BAT AGN show significant variability, so if the 0.4--10~keV data for a particular source was taken during the time-frame of the BAT survey, V13 renormalized the BAT spectrum based on the source flux in the BAT band at the observation time of the XMM/XRT data, using these renormalized spectra to determine $N_{\rm H}$. For this paper we use the uncorrected (non-renormalized) BAT spectra providing a consistent approach for all sources, since renormalization was only possible for 40\% of objects. The summed emission should anyhow be relatively independent of intra-source variability.

The stacked model fits below 10~keV do not have error estimates on the summed fluxes in each energy bin, but above 10~keV, for each energy channel we add the errors from each AGN in quadrature and take the square root of the sum to determine the total error. The total summed BAT spectrum has very small formal errors, but some of the $N_{\rm H}$ selected subsets have larger errors, particularly Compton-thick sources, because of the small number of such sources (12). Thus the Compton-thick fraction remains a key uncertainty in generating the total AGN spectrum, even when using a `complete' AGN  sample.  We caution that even hard X-ray selected catalogs are somewhat biased against Compton-thick AGN (\citealt{2009MNRAS.399..944M,2011ApJ...728...58B},V13).

The highest energy BAT channel has relatively large errors for all $N_{\rm H}$ groups perhaps indicative of a calibration problem in this channel; other studies (e.g., \citealt{2011ApJ...728...58B}) find similar anomalies. Our analysis is therefore most reliable below $\sim$150~keV, still covering the peak of the XRB emission, critical for determining the total AGN power budget.

We overplot the XRB spectrum from G07 (a composite from many missions), scaled down in flux to match the summed emission from the BAT sources in 2--200~keV. We employ the XRB spectral shape from HEAO-1 \citep{1987IAUS..124..611B} to determine the 2--200~keV `average' rest-frame XRB intensity (i.e. 1--100~keV for the observed XRB). Once correcting for our survey area (1.47 steradians), our stacked spectrum normalisation represents $1.2 \%$ of the XRB flux (i.e. 82 times lower in flux).   We cross-check this against the more recent measurement of the XRB using \emph{Swift}/XRT and BAT by \cite{2009A&A...493..501M} and find very similar results (our BAT sources constitute $\sim 1 \%$ of their XRB intensity).  Other determinations of the XRB using \emph{BeppoSAX} \citep{2007ApJ...666...86F} and \emph{INTEGRAL} \citep{2007A&A...467..529C,2010A&A...512A..49T} yield different overall normalizations, but the comparison between these and many other XRB measurements in \cite{2008ApJ...689..666A} (Figures 13 and 14) show consistency within errors, and we are mainly concerned with the spectral shape rather than the normalization in this study.  We also shift  the XRB spectrum into the `average' rest-frame, assuming an average redshift of XRB sources of $\sim$1.0. We also apply the normalization and redshift scalings to the XRB synthesis model components (for different $N_{\rm H}$ bins) in the G07 model.  

\section{Results}

The peak in the summed BAT spectrum coincides with that from the XRB at $\sim$60 keV (corresponding to $\sim$30~keV for $z\sim 1$ AGN). If the 0.4--200~keV spectral shape of AGN at $z\sim 0$ is representative of the AGN spectral shape throughout AGN history, this provides an elegant confirmation that most of the accretion contributing to the XRB occurs at $z \sim 1$.  The normalization of the $>10$~keV stacked spectrum lies above the scaled XRB emission; but the small number of sources (95) yields an intrinsic 10\% statistical fluctuation in our normalization, as does the uncertainty in the XRB normalization; the results are therefore consistent within errors. We test that our stacked spectrum is not dominated by a few bright sources by checking the effect of omitting the brightest three; the $>10$~keV spectral shape remains unchanged whereas the $<10$~keV spectral shape becomes moderately softer within the error range of the measured XRB slope.

The spectral evolution of AGN above 10~keV (rest-frame) is not known, but a key result of our study is that the combined emission from local AGN is consistent with the (redshift-corrected) XRB shape. Below 10~keV, the contribution from unobscured AGN lies a factor of $\sim 1.4$ lower than the G07 model synthesis prediction while that from obscured Compton-thin AGN agrees well with the synthesis model.  Above 10~keV, the unobscured AGN contribution agrees reasonably well; however the obscured Compton-thin AGN contribution lies $\sim 78\%$ higher but with a similar shape to the G07 model. 

\begin{figure*}
\centerline{
\includegraphics[width=9.0cm]{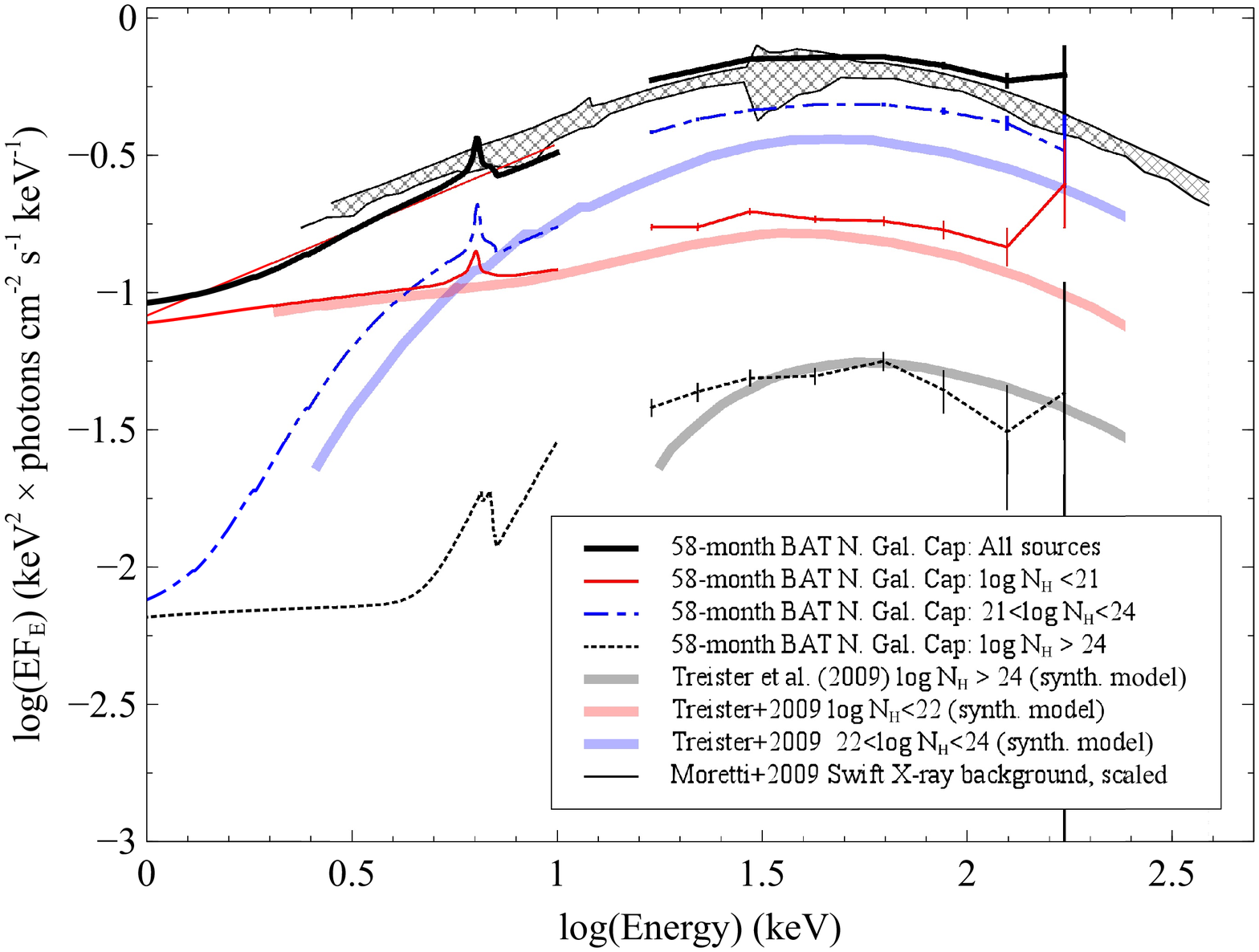}\\\\
\includegraphics[width=9.0cm]{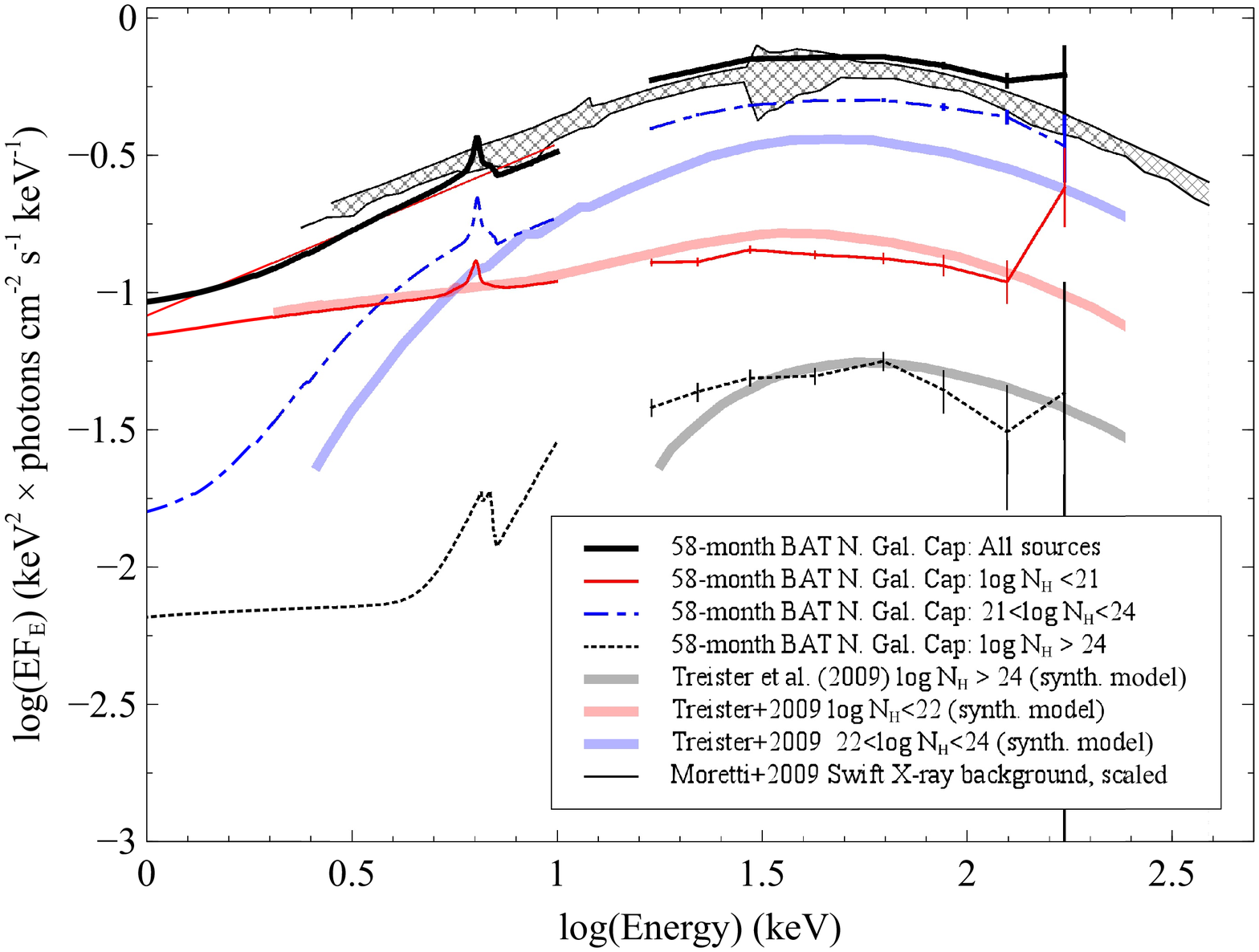}
}
\caption{Comparison with the T09 XRB synthesis model. Key as in Fig.~\ref{stackedspectra}, except the boundary between unabsorbed and absorbed groups is now $\rm log N_{\rm H} = 22$, as employed by T09.\label{stackedspectra_Treister09comp}}
\end{figure*}

We also compare our results with the T09 XRB synthesis model in Fig.~\ref{stackedspectra_Treister09comp}, splitting obscured and unobscured sources at $\rm log \thinspace N_{\rm H} = 22$ instead, as done by those authors. For down-scaling the XRB shape and model components, here we employ the more recent \emph{Swift} XRB measurement by \cite{2009A&A...493..501M}, which has less overall uncertainty in the normalization. Our unobscured and Compton-thick spectra show good agreement, but the obscured class shows a higher contribution both below and above 10~keV, with an excess above 10~keV consistent with stronger reflection, perhaps indicating that Compton reflection in Compton thin sources is more important than previously assumed.

The Compton-thick contribution above 10~keV lies below the predictions from the G07 model and is consistent with the T09 model; the higher hard X-ray excess in Compton-thin obscured sources requires less of a Compton-thick component to produce the overall XRB shape. 

\cite{2012A&A...546A..98A} find that reflection at the level of $0.5<R<2$ can be combined with the observed Compton-thick fraction from \emph{Swift} studies (e.g. \citealt{2011ApJ...728...58B}) to explain the XRB. The offset seen here may be too great to be from simple reflection geometries and might require strong light bending for a substantial fraction of sources (e.g. \citealt{2007MNRAS.382.1005G}).  Such sources represent over 50$\%$ of the sample, at the upper limit of the proportion of light-bent sources predicted \cite{2007MNRAS.382.1005G} scenario.  A substantial fraction of the individual sources (V13 and \citealt{2013ApJ...762...80T}) show such a `hard excess' consistent with either a higher than expected reflection or an additional highly absorbed spectral component; the other absorption groups do not show such an excess. 

\subsection{Parameterising the hard excess using reflection}

\cite{2011A&A...532A.102R} analyze the INTEGRAL spectra ($>10$~keV) of 165 hard X-ray selected AGN and perform a comprehensive and complementary analysis. They find that moderately obscured AGN (`MOB', $23 < \rm log N_{\rm H} < 24$) have more pronounced reflection ($R = 2.2^{+4.5}_{-1.1}$) than unabsorbed ($\rm log N_{\rm H} < 22$) and lightly-obscured AGN (`LOB', $\rm 22 < log N_{\rm H} < 23$). 

We split our sample into their categories (39 unabsorbed, 18 LOB, 25 MOB, 12 Compton-thick, {assuming minimal $N_{\rm H}$ when two are available). For the averaged model fits $<10$~keV, we sample the spectra with 20 bins below 10~keV, assume a conservative value of $10 \%$ errors on the flux density in each bin (if the data themselves were stacked, errors on the total spectrum would be much smaller; {we estimate errors on the Compton-thick group to be 6\%, which should represent the worst case). We fit the 1--200~keV data with the reflection model {\sc pexrav}, assuming defaults for all parameters, allowing the photon index $\Gamma$ and reflection fraction $R$ to vary (we freeze $E_{\rm fold}$ at 300~keV consistent with the average for BAT AGN, see V13).  We assume the same normalization below and above 10~keV, since intra-object variability should be smoothed out in a stacked spectrum, and any statistically significant difference in normalization between soft and hard bands should therefore be real.  We include a partial covering absorber for sources with $\rm log N_{\rm H}>22$ (model {\sc pcfabs(pexrav)}), and find that the $N_{\rm H}$ value obtained from the stacked spectra fits are well within the $N_{\rm H}$ limits of each bin.   

Our results (Fig.~\ref{refl_vs_gamma_cont}) are in good agreement with \cite{2011A&A...532A.102R}, finding a pronounced reflection component for MOB AGN (R=$2.0_{-0.5}^{+0.6}$).  One possible explanation \citep{2011A&A...532A.102R} is that for moderately-obscured sources, reflection from Compton-thick clumpy clouds of absorbing material boosts the total reflected flux. The {\sc pexrav} model produces a worse fit to MOB AGN ($\chi^{2} / \rm d.o.f. \approx 8$ compared to $\sim 1$ for other classes) overpredicting the 5--10~keV emission, which can be partially resolved by allowing for sub-solar abundances in the {\sc pexrav} fit; however it is unclear why MOB AGN should have intrinsically different abundances.  The contours show an evolution towards higher reflection fraction with higher absorption.  We assumed a constant inclination angle; in the standard Unified Model \citep{1993ARA&A..31..473A} one may expect more edge-on tori for more heavily-absorbed sources.  We investigated this effect along the same lines as \cite{2011A&A...532A.102R} and find that if inclination increases with absorption, the reflection fraction increases even more strongly with absorption than for a constant angle, indicating a salient difference in the reflection geometry.  The reflection fraction for Compton-thick AGN is more uncertain and requires more sophisticated modeling, so we omit it from Fig.~\ref{refl_vs_gamma_cont}.

The difference of photon indices between absorbed and unabsorbed sources ($\Gamma = 1.86 \pm 0.05$ for unabsorbed and $\Gamma = 1.76 \pm 0.03$ for MOB AGN) is very similar to that seen by \cite{2013arXiv1302.1062L} for the COSMOS field ($\Gamma=1.89 \pm 0.02$ for Seyfert 1s and $\Gamma = 1.76 \pm 0.03$ for Seyfert 2s). In the context of a Compton-scattering model for the X-ray continuum, it may indicate that the optical depth `seen' by accretion disk photons emerging from the Comptonizing corona varies depending on the angle at which the corona is viewed. This indicates some degree of deviation from spherical geometry for the corona, with implications for AGN unification schemes. 

One can potentially also account for these results if luminosity is the key driver: since lower luminosity sources have a greater Iron K$\alpha$ linewidth \citep{1993ApJ...413L..15I} and thus an implied higher reflection (\citealt{1997ApJ...488L..91N}, \citealt{2005A&A...432...15P}), the observed changes in reflection with absorption can be ascribed to a luminosity effect rather than geometry.  However, stacking the unabsorbed AGN in four different luminosity bins with equal numbers of objects per bin (Fig.~\ref{refl_vs_lumin_unabs}), shows no trend for luminosity-dependent reflection (for $\rm log \thinspace L_{\rm 2-10 keV} < 44.4$). This remains a debated issue: such a trend has been observed to varying degrees in smaller samples of high-luminosity quasars \citep{1992ApJ...389..157W,2005MNRAS.364..195P}, but our sample does not extend to such luminosities to test this.

\begin{figure}
\centerline{
\includegraphics[width=9.0cm]{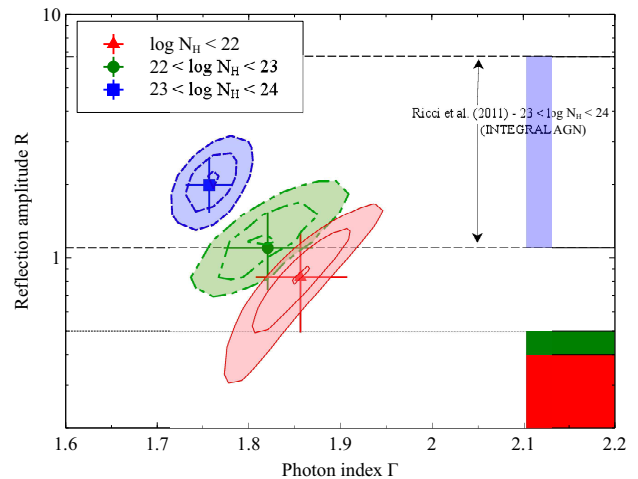}
}
\caption{Contours for reflection amplitude $R$ against photon index $\Gamma$.  Red solid line contours: `unabsorbed' ($\rm log \thinspace N_{\rm H} \thinspace < 22$) AGN; green dot-dashed contours: `LOB' AGN ($22 < \rm log \thinspace N_{\rm H} \thinspace < 23$); blue dashed contours: `MOB' AGN ($23 < \rm log \thinspace N_{\rm H} \thinspace < 24$).  Contours represent a total change in $\chi^{2}$ of 10.0 from the minimal $\chi^{2}$ (number of degrees of freedom are 15, 13 and 15 for unabsorbed, LOB and MOB classes respectively).  Colored shaded regions on the right-hand side of the plot indicate the ranges of reflection parameters found for different absorption groups by \cite{2011A&A...532A.102R}. The position of these shaded areas does not reflect the photon index distribution from \cite{2011A&A...532A.102R}. \label{refl_vs_gamma_cont}}
\end{figure}

\begin{figure}
\centerline{
\includegraphics[width=9.0cm]{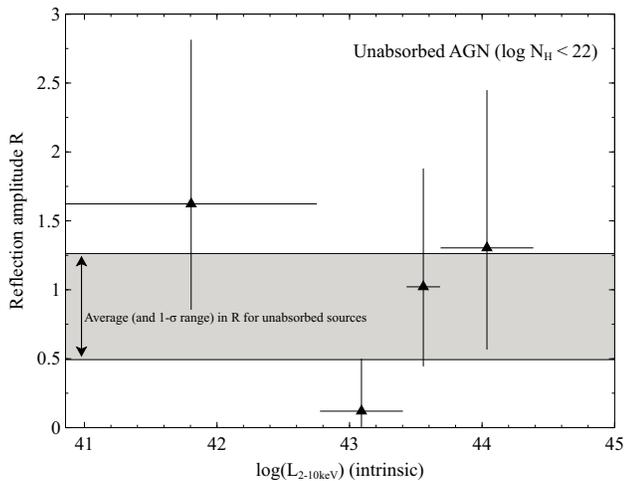}
}
\caption{Reflection vs intrinsic 2--10~keV luminosity for $\rm log \thinspace N_{\rm H} < 22$ sources, stacked in different luminosity bins. \label{refl_vs_lumin_unabs}}
\end{figure}

\subsection{Objects without 0.1--10~keV spectra and their contribution above 10~keV}

Five objects in V13 have poor quality 0.4--10~keV data, precluding accurate determinations of $N_{\rm H}$; they have not been included in the total stacked spectrum. Since these sources contribute only $\sim 3\%$ to the total emission at 30~keV, their net effect is small (Fig.~\ref{stackedspectra_uplimonly}). In the extreme case, if they are all Compton-thick, the percentage of Compton-thick sources could be as high as $\sim 18 \%$.

\begin{figure}
\centerline{
\includegraphics[width=9.0cm]{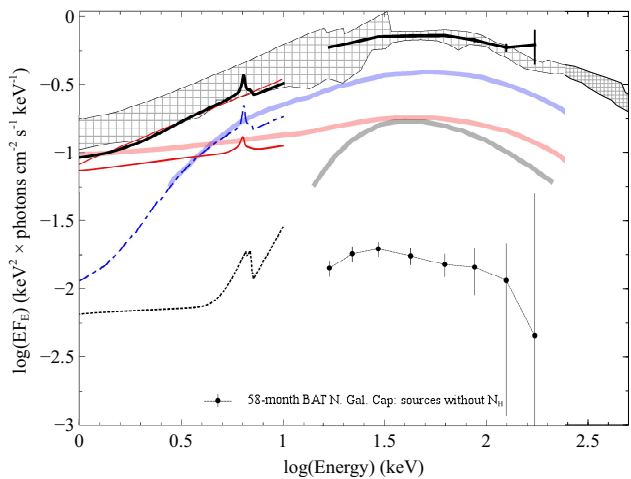}\\
}
\caption{Key as for Fig.~\ref{stackedspectra}.  Filled circles joined by dotted lines represent the contribution from five sources without sufficient counts at 0.4--10~keV to determine their $N_{\rm H}$ values and contribution below 10~keV.\label{stackedspectra_uplimonly}}
\end{figure}

\section{Summary}

We present the summed 0.4--200~keV spectrum of the AGN in Northern Galactic Cap ($b > 50^{\circ}$) of the 58-month BAT catalog, and compare the observed contribution of AGN with different absorptions to predictions from XRB synthesis models.  There is a striking similarity between our stacked spectrum and the XRB spectral shape.  Despite the known evolution in the luminosity function with redshift, the total emission from sources at $z \sim 0$ is very similar to the sum over all redshifts from the XRB. Errors in the XRB normalization and Poisson fluctuations in our source numbers contribute uncertainties of $\sim 10 \%$ to the relative normalization.

This low-redshift analysis has the considerable benefit of good-quality data across the entire 0.4--200 keV band, unavailable for much fainter higher redshift sources that make up the bulk of the XRB, allowing a direct decomposition into component source populations across a substantial part of the XRB spectral energy range. Most importantly, our analysis covers the peak emission at $\sim 30$~keV.   We employ a complete, relatively absorption-unbiased subsample drawn from a hard X-ray selected catalog.

Comparing the stacked spectra to the G07 synthesis model, below 10~keV we find excellent agreement in the spectral shapes and the proportions of sources required to make up the emission. Above 10~keV: 1) the total emission has a slightly higher normalization than the XRB but within errors; 2) the hard excess or effective `reflection' component for absorbed Compton-thin sources is higher than predicted by G07; and 3) the Compton-thick contribution is between 20--40$\%$ lower than the G07 model prediction, broadly consistent with the T09 model.  Compton reflection, or an equivalent process, contributes to the hard excess particularly for $23 < \rm log N_{\rm H} < 24$ sources, as found by \cite{2011A&A...532A.102R} for INTEGRAL AGN.  This does not seem to be due to an anti-correlation of luminosity with reflection, which is absent in our sample.

Our observed Compton-thick fraction is consistent with the G07 prediction \emph{at our survey flux limit} (Fig.~17 of G07), even though it is not consistent with the overall flux- and luminosity-integrated Compton-thick fraction of the G07 model.  However, we consistently compare our results with the full flux- and luminosity-integrated model synthesis results, since the key point here is that the stacked emission of a local, bright, but proportionately tiny fraction of the XRB surprisingly reproduces its shape, \emph{despite} the flux limit.  As such, these findings are not necessarily extendable to the entire XRB synthesis-model parameter space, but suggest interesting possibilites for future studies.

We have not included the contribution of blazars and other jet-dominated objects but they may be a significant contributor to the XRB; \cite{2009ApJ...707..778D} find a 7$\%$ contribution in 2--10~keV and 9$\%$ in 15--55~keV.  However, blazars are only 3$\%$ of our $z<0.2$ sample, and are therefore not significant here.

The integrated spectrum at $z = 0$ is surprisingly similar to that averaged over all cosmic time despite a known strong evolution in properties with redshift (\citealt{2010MNRAS.401.2531A}, \citealt{2003ApJ...598..886U}). There is accumulating evidence now that the XRB shape is consistent with a smaller Compton-thick fraction (13\%) and higher reflection (notably $R=2.0$ for moderately absorbed $23 < \rm log \thinspace N_{\rm H} \thinspace < 24$ sources), supporting the idea that such sources may be the main contributor to the XRB \citep{1990MNRAS.242P..14F,2007MNRAS.382.1005G}, or that very complex absorption may play a role \citep{2013ApJ...762...80T}. \emph{NuSTAR} observations will be key for definitively answering these questions.

\section{Acknowledgements}

We thank the anonymous referee, Prof. R. Gilli and Dr. R. Krivonos for helpful suggestions which improved this work.  We thank Dr. A. Akylas and Dr. A. Georgakakis for clarifying some of the results in \cite{2012A&A...546A..98A} and Prof. W. N. Brandt for useful discussions on the 58-month BAT catalog Northern Galactic Cap sample.

\bibliographystyle{apj} 
\bibliography{stacked58monthBATspec}

\end{document}